# An Introduction to MM Algorithms for Machine Learning and Statistical Estimation

Hien D. Nguyen

November 12, 2016

School of Mathematics and Physics, University of Queensland, St. Lucia.

Centre for Advanced Imaging, University of Queensland, St. Lucia.

## Abstract

MM (majorization–minimization) algorithms are an increasingly popular tool for solving optimization problems in machine learning and statistical estimation. This article introduces the MM algorithm framework in general and via three popular example applications: Gaussian mixture regressions, multinomial logistic regressions, and support vector machines. Specific algorithms for the three examples are derived and numerical demonstrations are presented. Theoretical and practical aspects of MM algorithm design are discussed.

## 1 Introduction

Let $\boldsymbol{X} \in \mathbb{X} \subset \mathbb{R}^p$ and $\boldsymbol{Y} \in \mathbb{Y} \subset \mathbb{R}^q$ be random variables, which we shall refer to as the input and target variables, respectively. We shall denote a sample of $n$ independent and identically distributed (IID) pairs of variables $\mathbb{D} = \{(\boldsymbol{X}_i, \boldsymbol{Y}_i)\}_{i=1}^n$ as the data, and $\bar{\mathbb{D}} = \{(\boldsymbol{x}_i, \boldsymbol{y}_i)\}_{i=1}^n$ as an observed realization of the data. Under the empirical risk minimization (ERM) framework of Vapnik (1998, Ch. 1) or the extremum estimation (EE) framework of Amemiya (1985, Ch. 4), a large number of machine learning and statistical estimation problems can be phrased as the computation of

$$\min_{\boldsymbol{\theta} \in \Theta} \mathcal{R}\left(\boldsymbol{\theta}; \bar{\mathbb{D}}\right) \text{ or } \hat{\boldsymbol{\theta}} = \arg\min_{\boldsymbol{\theta} \in \Theta} \mathcal{R}\left(\boldsymbol{\theta}; \bar{\mathbb{D}}\right), \qquad (1)$$

where $\mathcal{R}\left(\boldsymbol{\theta}; \bar{\mathbb{D}}\right)$ is a risk function defined over the observed data $\bar{\mathbb{D}}$ and is dependent on some parameter $\boldsymbol{\theta} \in \Theta$.

Common risk functions that are used in practice are the negative log-likelihood functions, which can be expressed as

$$\mathcal{R}\left(\boldsymbol{\theta}; \bar{\mathbb{D}}\right) = -\frac{1}{n} \sum_{i=1}^n \log f\left(\boldsymbol{x}_i, \boldsymbol{y}_i; \boldsymbol{\theta}\right),$$

where $f\left(\boldsymbol{x}, \boldsymbol{y}; \boldsymbol{\theta}\right)$ is a density function over the support of $\boldsymbol{X}$ and $\boldsymbol{Y}$, which takes parameter $\boldsymbol{\theta}$. The minimization of the risk in this case yields the maximum likelihood (ML) estimate for the data $\bar{\mathbb{D}}$, given the parametric family $f$. Another common risk function is the $|\cdot|^d$-norm difference between the target variable and some function $f$ of the input:

$$\mathcal{R}\left(\boldsymbol{\theta}; \bar{\mathbb{D}}\right) = \frac{1}{n} \sum_{i=1}^n |y_i - f\left(\boldsymbol{x}_i; \boldsymbol{\theta}\right)|^d,$$

where $d \in [1, 2]$, $y_i \in \mathbb{R}$, and $f\left(\boldsymbol{x}; \boldsymbol{\theta}\right)$ is some predictive function of $y_i$ with parameter $\boldsymbol{\theta}$ that takes $\boldsymbol{x}$ as an input. Setting $d = 1$ and $d = 2$ yield the common least-absolute deviation and least-squares criteria, respectively. Furthermore, taking $f\left(\boldsymbol{x}; \boldsymbol{\theta}\right) = \boldsymbol{\theta}^\top \boldsymbol{x}$ and $d = 2$ simultaneously yields the classical ordinary least-squares criterion. Here $\Theta \subset \mathbb{R}^p$ and the superscript $\top$ indicates matrix transposition.

When $\mathbb{Y} = \{-1, 1\}$, a common problem in machine learning is to construct a classification function $f(\boldsymbol{x}_i; \boldsymbol{\theta})$ that minimizes the classification (0-1) risk

$$\mathcal{R}(\boldsymbol{\theta}; \bar{\mathbb{D}}) = \frac{1}{n} \sum_{i=1}^{n} \mathbb{I}\{y_i \neq f(\boldsymbol{x}_i; \boldsymbol{\theta})\},$$

where $f: \mathbb{X} \to \mathbb{Y}$ and $\mathbb{I}\{A\} = 1$ if proposition $A$ is true and 0 otherwise. Unfortunately, the form of the classification risk is combinatorial and thus necessitates the use of surrogate classification risks of the form

$$\mathcal{R}(\boldsymbol{\theta}; \bar{\mathbb{D}}) = \frac{1}{n} \sum_{i=1}^{n} \psi(\boldsymbol{x}_i, y_i, f(\boldsymbol{x}_i; \boldsymbol{\theta})),$$

where $\psi: \mathbb{R}^p \times \{-1, 1\}^2 \to [0, \infty)$ and $\psi(\boldsymbol{x}, y, y) = 0$ for all $\boldsymbol{x}$ and $y$ [cf. Scholkopf & Smola (2002, Def. 3.1)]. An example of a machine learning algorithm that minimizes a surrogate classification risk is the support vector machine (SVM) of Cortes & Vapnik (1995). The linear-basis SVM utilizes a surrogate risk function, where $\psi(\boldsymbol{x}, y, f(\boldsymbol{x}; \boldsymbol{\theta})) = \max\{0, 1 - yf(\boldsymbol{x}; \boldsymbol{\theta})\}$ is the hinge loss function, $f(\boldsymbol{x}; \boldsymbol{\theta}) = \alpha + \boldsymbol{\beta}^\top \boldsymbol{x}$, and $\boldsymbol{\theta}^\top = (\alpha, \boldsymbol{\beta}^\top) \in \Theta \subset \mathbb{R}^{p+1}$.

The task of computing (1) may be complicated by various factors that fall outside the scope of the traditional calculus formulation for optimization [cf. Khuri (2003, Ch. 7)]. Such factors include the lack of differentiability of $\mathcal{R}$ or difficulty in obtaining closed-form solutions to the first-order condition (FOC) equation $\nabla_{\boldsymbol{\theta}} \mathcal{R} = \boldsymbol{0}$, where $\nabla_{\boldsymbol{\theta}}$ is the gradient operator with respect to $\boldsymbol{\theta}$, and $\boldsymbol{0}$ is a zero vector.

The MM (majorization–minimization) algorithm framework is a unifying paradigm for simplifying the computation of (1) when difficulties arise, via iterative minimization of surrogate functions. MM algorithms are particularly attractive due to the monotonicity and thus stability of their objective sequences as well as global convergence of their limits, in general settings.

A comprehensive treatment on the theory and implementation of MM algorithms can be found in Lange (2016). Summaries and tutorials on MM algorithms for various problems can be found in Becker et al. (1997), Hunter & Lange (2004), Lange (2013, Ch. 8), Lange et al. (2000), Lange et al. (2014), McLachlan & Krishnan (2008, Sec. 7.7), Wu & Lange (2010), and Zhou et al. (2010). Some theoretical analyses of MM algorithms can be found in de Leeuw & Lange (2009), Lange (2013, Sec. 12.4), Mairal (2015), and Vaida (2005).

It is known that MM algorithms are generalizations of the EM (expectation–maximization) algorithms of Dempster et al. (1977) [cf. Lange (2013, Ch. 9)]. The recently established connection between MM algorithms and the successive upper-bound maximization (SUM) algorithms of Razaviyayn et al. (2013) further shows that the MM algorithm framework also covers the concave-convex procedures [Yuille & Rangarajan (2003); CCCP], proximal algorithms (Parikh & Boyd, 2013), forward-backward splitting algorithms [Combettes & Pesquet (2011); FBS], as well as various incarnations of iteratively-reweighed least-squares algorithms (IRLS) such as those of Becker et al. (1997) and Lange et al. (2000); see Hong et al. (2016) for details.

It is not possible to provide a complete list of applications of MM algorithms to machine learning, statistical estimation, and signal processing problems. We present a comprehensive albeit incomplete summary of applications of MM algorithms in Table 1. Further examples and references can be found in Hong et al. (2016) and Lange (2016).

In this article, we will present the MM algorithm framework via applications to three examples that span the scope of statistical estimation and machine learning problems: Gaussian mixtures of regressions (GMR), multinomial-logistic regressions (MLR), and SVM estimations. The three estimation prob-lems will firstly be presented in Section 2. The MM algorithm framework will be presented in Section 3 along with some theoretical results. MM algorithms for the three estimation problems are presented in Section 4. Numerical demonstrations of the MM algorithms are presented in Section 5. Conclusions are then drawn in Section 6.

Table 1: MM algorithm applications and references.

| Application | References |
| --- | --- |
| Bradley-Terry models estimation | (Hunter, 2004; Lange et al., 2000) |
| Convex and shape-restricted regressions | (Chi et al., 2014) |
| Dirichlet-multinomial distributions estimation | (Zhou & Lange, 2010; Zhou & Zhang, 2012) |
| Elliptical symmetric distributions estimation | (Becker et al., 1997) |
| Fully-visible Boltzmann machines estimation | (Nguyen & Wood, 2016) |
| Gaussian mixtures estimation | (Nguyen & McLachlan, 2015) |
| Geometric and sigmoidal programming | (Lange & Zhou, 2014) |
| Heteroscedastic regressions | (Daye et al., 2012; Nguyen et al., 2016a) |
| Laplace regression models estimation | (Nguyen & McLachlan, 2016a; Nguyen et al., 2016b) |
| Least $|\cdot|^d$-norm regressions | (Becker et al., 1997; Lange et al., 2000) |
| Linear mixed models estimation | (Lange et al., 2014) |
| Logistic and multinomial regressions | (Bohning & Lindsay, 1988; Bohning, 1992) |
| Markov random field estimation | (Nguyen et al., 2016c) |
| Matrix completion and imputation | (Mazumder et al., 2010; Lange et al., 2014) |
| Mixture of experts models estimation | (Nguyen & McLachlan, 2014, 2016a) |
| Multidimensional scaling | (Lange et al., 2000) |
| Multivariate $t$ distributions estimation | (Wu & Lange, 2010) |
| Non-negative matrix factorization | (Lee & Seung, 1999) |
| Poisson regression estimation | (Lange et al., 2000) |
| Point to set minimization problems | (Chi & Lange, 2014; Chi et al., 2014) |
| Polygonal distributions estimation | (Nguyen & McLachlan, 2016b) |
| Quantile regression estimation | (Hunter & Lange, 2000) |
| SVM estimation | (de Leeuw & Lange, 2009; Wu & Lange, 2010) |
| Transmission tomography image reconstruction | (De Pierro, 1993; Becker et al., 1997) |
| Variable selection in regression via regularization | (Hunter & Li, 2005; Lange et al., 2014) |

## 2 Example problems

### 2.1 Gaussian mixture of regressions

Let $\boldsymbol{X}$ arise from a distribution with unknown density function $f_X(\boldsymbol{x})$, which does not depend on the parameter $\boldsymbol{\theta}$ ($\boldsymbol{X}$ can be non-stochastic). Conditioned on $\boldsymbol{X} = \boldsymbol{x}$, suppose that $\boldsymbol{Y}$ can arise from one of $g \in \mathbb{N}$ possible component regression regimes. Let $Z$ be a random variable that indicates the component from which $\boldsymbol{Y}$ arises, such that $\mathbb{P}(Z = c) = \pi_c$, where $c \in [g]$ ($[g] = \{1, 2, ..., g\}$), $\pi_c > 0$, and $\sum_{c=1}^{g} \pi_c = 1$. Write the conditional probability density of $\boldsymbol{Y}$ given $\boldsymbol{X} = \boldsymbol{x}$ and $Z = c$ as

$$f_{Y|X,c}(\boldsymbol{y}|\boldsymbol{x}; \mathbf{B}_c, \boldsymbol{\Sigma}_c) = \phi(\boldsymbol{y}; \mathbf{B}_c \boldsymbol{x}, \boldsymbol{\Sigma}_c), \qquad (2)$$

where $\mathbf{B}_c \in \mathbb{R}^{q \times p}$, $\boldsymbol{\Sigma}_c$ is a positive-definite $q \times q$ matrix covariance matrix, and

$$\phi(\boldsymbol{y}; \boldsymbol{\mu}, \boldsymbol{\Sigma}) = (2\pi)^{-d/2} |\boldsymbol{\Sigma}|^{-1/2} \exp\left[-\frac{1}{2}(\boldsymbol{y} - \boldsymbol{\mu})^\top \boldsymbol{\Sigma}(\boldsymbol{y} - \boldsymbol{\mu})\right] \qquad (3)$$

is the multivariate Gaussian distribution with mean vector $\boldsymbol{\mu}$ and covariance matrix $\boldsymbol{\Sigma}$. The conditional (in $Z$) characterization (2) leads to the marginal characterization

$$f_{Y|X}(\boldsymbol{y}|\boldsymbol{x}; \boldsymbol{\theta}) = \sum_{c=1}^{g} \pi_c \phi(\boldsymbol{y}; \mathbf{B}_c \boldsymbol{x}, \boldsymbol{\Sigma}_c), \qquad (4)$$

where $\boldsymbol{\theta}$ contains the parameter elements $\pi_c$, $\mathbf{B}_c$, and $\boldsymbol{\Sigma}_c$, for $c \in [g]$. We refer to the characterization (4) as the GMR model.

The GMR model was first proposed by Quandt (1972) for the $q = 1$ case and an EM algorithm for the same case was proposed in DeSarbo & Cron (1988). To the best of our knowledge, the general multivariate case ($q > 1$) of characterization 4 was first considered in Jones & McLachlan (1992). See McLachlan & Peel (2000) regarding mixture models in general.

Given data $\bar{\mathbb{D}}$, the estimation of a GMR model requires the minimization of the negative log-likelihood risk

$$\begin{aligned} \mathcal{R}\left(\boldsymbol{\theta}; \bar{\mathbb{D}}\right) &= -\frac{1}{n} \sum_{i=1}^{n} \log f_{Y|X}\left(\boldsymbol{y}_i | \boldsymbol{x}_i; \boldsymbol{\theta}\right) \\ &= -\frac{1}{n} \sum_{i=1}^{n} \log \sum_{c=1}^{g} \pi_c \phi\left(\boldsymbol{y}_i; \mathbf{B}_c \boldsymbol{x}_i, \boldsymbol{\Sigma}_c\right). \end{aligned} \quad (5)$$

The difficulty with computing (1) for (5) arises due to the lack of a closed-form solution to the FOC equation $\nabla_{\boldsymbol{\theta}} \mathcal{R} = \mathbf{0}$. This is due to the log-sum-exp functional form that is embedded in each log-likelihood element.

## 2.2 Multinomial-logistic regressions

Let $\boldsymbol{X}$ arise from a distribution with unknown density function $f_X(\boldsymbol{x})$, which does not depend on the parameter $\boldsymbol{\theta}$ ($\boldsymbol{X}$ can be non-stochastic). Suppose that $\mathbb{Y} = [g]$ for $g \in \mathbb{N}$ and let the conditional relationship between $Y$ and $\boldsymbol{X}$ be characterized by

$$\mathbb{P}\left(Y = c | \boldsymbol{X} = \boldsymbol{x}; \boldsymbol{\theta}\right) = \frac{\exp\left(\boldsymbol{\beta}_c^\top \boldsymbol{x}\right)}{\sum_{d=1}^{g} \exp\left(\boldsymbol{\beta}_d^\top \boldsymbol{x}\right)}, \quad (6)$$

where $\boldsymbol{\theta}$ contains the parameter elements $\boldsymbol{\beta}_c \in \mathbb{R}^p$ for $c \in [g-1]$, and $\boldsymbol{\beta}_g = \mathbf{0}$. We refer to the characterization (2.2) as the MLR model.

The MLR model is a well-studied and widely applied formulation for categorical variable regression in practice. See for example Amemiya (1985, Sec. 9.3) and Greene (2003, Sec. 21.7.1) for a statistical and econometric perspective, and Bishop (2006, Sec. 4.3.4) for a McLachlan (1992, Ch. 8) for some machine learning and pattern recognition points of view.

Given data $\bar{\mathbb{D}}$, the estimation of a MLR model requires the minimization of the negative log-likelihood risk

$$\begin{aligned}
\mathcal{R}\left(\boldsymbol{\theta};\bar{\mathbb{D}}\right) &= -\frac{1}{n}\sum_{i=1}^{n}\prod_{c=1}^{g}\left[\mathbb{P}\left(Y=c|\boldsymbol{X}=\boldsymbol{x};\boldsymbol{\theta}\right)\right]^{\mathbb{I}\{y=c\}} \\
&= -\frac{1}{n}\sum_{c=1}^{g}\sum_{i=1}^{n}\mathbb{I}\{y_i=c\}\boldsymbol{\beta}_c^\top \boldsymbol{x}_i + \frac{1}{n}\sum_{i=1}^{n}\log\sum_{c=1}^{g}\exp\left(\boldsymbol{\beta}_c^\top \boldsymbol{x}_i\right). \quad (7)
\end{aligned}$$

The difficulty with computing (1) for (7), like (5), arises from the lack of a closed form solution to the FOC equation $\nabla_{\boldsymbol{\theta}}\mathcal{R} = \mathbf{0}$. Due to the general convexity of MLR risk [cf. Albert & Anderson (1984)], the usual strategy for computing (1) is via a Newton-Raphson algorithm.

Let $\mathcal{H}_{\boldsymbol{\theta}} = \nabla_{\boldsymbol{\theta}}\nabla_{\boldsymbol{\theta}^\top}$ be the Hessian operator. It is noted in Bishop (2006, Sec. 4.3.4) that the Hessian $\mathcal{H}_{\boldsymbol{\theta}}\mathcal{R}$ consists of $(g-1)g/2$ blocks of $p \times p$ matrices with forms

$$\nabla_{\boldsymbol{\beta}_c}\nabla_{\boldsymbol{\beta}_d^\top}\mathcal{R} = \sum_{i=1}^{n}\mathbb{I}\{y_i=c\}\left(\mathbf{I}_{[c,d]} - \mathbb{I}\{y_i=c\}\right)\boldsymbol{x}_i\boldsymbol{x}_i^\top,$$

for $c,d \in [g-1]$ and $c \leq d$, where $\mathbf{I}$ is the identity matrix and $\mathbf{I}_{[c,d]}$ is the element in its $c$th row and $d$th column. Since $\mathcal{H}_{\boldsymbol{\theta}}\mathcal{R}$ is therefore $[(g-1)p] \times [(g-1)p]$, inversion may be difficult for large values of either $g$ or $p$. Thus, a method that avoids the full computation or inversion of the Hessian is desirable.

### 2.3 Support vector machines

Let $\boldsymbol{X}$ arise from a distribution with unknown density function $f_X(\boldsymbol{x})$, which does not depend on the parameter $\boldsymbol{\theta}$ ($\boldsymbol{X}$ can be non-stochastic). Suppose that $\mathbb{Y} = \{-1, 1\}$ and the relationship between $\boldsymbol{X}$ and $Y$ is unknown. For a linear-basis SVM, we wish to construct an optimal hyperplane (i.e. $\alpha + \boldsymbol{\beta}^\top \boldsymbol{X} = 0$; $\alpha \in \mathbb{R}$ and $\boldsymbol{\beta} \in \mathbb{R}^p$) such that the signs of $Y$ and $\alpha + \boldsymbol{\beta}^\top \boldsymbol{X}$ are the same, with high probability. From data $\bar{\mathbb{D}}$, an optimal hyperplane can be estimated by computing (1) for the risk

$$\mathcal{R}(\boldsymbol{\theta}; \bar{\mathbb{D}}) = \frac{1}{n} \sum_{i=1}^n \mathbb{I}\left\{y_i \neq \operatorname{sign}\left(\alpha + \boldsymbol{\beta}^\top \boldsymbol{x}_i\right)\right\}, \tag{8}$$

where $\operatorname{sign}(x) = 1$ if $x > 0$, $\operatorname{sign}(x) = -1$ if $x < 0$, and $\operatorname{sign}(0) = 0$. Here, $\boldsymbol{\theta}^\top = (\alpha, \boldsymbol{\beta}^\top) \in \Theta \subset \mathbb{R}^{p+1}$. Since (8) is combinatorial in nature, it is difficult to manipulate. As such, a surrogate risk function can be constructed from the hinge loss function $\psi(\boldsymbol{x}, y, f(\boldsymbol{x}; \boldsymbol{\theta})) = \max\left\{0, 1 - y\left(\alpha + \boldsymbol{\beta}^\top \boldsymbol{x}\right)\right\}$ to obtain the form

$$\begin{aligned}\mathcal{R}(\boldsymbol{\theta}; \bar{\mathbb{D}}) &= \frac{1}{n} \sum_{i=1}^n \psi(\boldsymbol{x}_i, y_i, f(\boldsymbol{x}_i; \boldsymbol{\theta})) \\ &= \frac{1}{n} \sum_{i=1}^n \max\left\{0, 1 - y_i\left(\alpha + \boldsymbol{\beta}^\top \boldsymbol{x}_i\right)\right\}.\end{aligned}$$

Finally, it is common practice to add a quadratic penalization term to avoid overfitting to the data and improve the generalizability of the estimated hyperplane. Under penalization, the linear-basis SVM can be estimated by computing (1) for the surrogate risk

$$\mathcal{R}(\boldsymbol{\theta}; \bar{\mathbb{D}}) = \frac{1}{n} \sum_{i=1}^n \max\left\{0, 1 - y_i\left(\alpha + \boldsymbol{\beta}^\top \boldsymbol{x}_i\right)\right\} + \lambda \boldsymbol{\beta}^\top \boldsymbol{\beta}, \tag{9}$$

where $\lambda \geq 0$ is a penalty term.

The difficulty in computing (1) for (9) arises due to the lack of differentiability of (9) in $\boldsymbol{\theta}$, due to the hinge loss function. Traditionally, (1) has been computed via a constrained quadratic programming formulation of (9) using Karush-Kuhn-Tucker (KKT) conditions; see Burges (1998) for example. We will demonstrate that it is possible to compute (1) without a constrained formulation via an MM algorithm.

Since the introduction of SVMs by Cortes & Vapnik (1995), there have been numerous articles and volumes written on the topic. Some high-quality texts in the area include Herbrich (2002), Scholkopf & Smola (2002), and Steinwart & Christmann (2008).

## 3 MM algorithms

Suppose that we wish to obtain

$$\min_{\boldsymbol{\theta}\in\Theta} \mathcal{O}(\boldsymbol{\theta}) \text{ or } \hat{\boldsymbol{\theta}} = \arg\min_{\boldsymbol{\theta}\in\Theta} \mathcal{O}(\boldsymbol{\theta}), \qquad (10)$$

for some difficult to manipulate objective function $\mathcal{O}$, where $\Theta$ is a subset of some Euclidean space. Instead of operating on $\mathcal{O}$, we can sconsider an easier to manipulate majorizer of $\mathcal{O}$ at some point $\boldsymbol{v} \in \Theta$ instead.

**Definition 1.** We say that $\mathcal{M}(\boldsymbol{\theta}; \boldsymbol{v})$ is a majorizer of objective $\mathcal{O}(\boldsymbol{\theta})$ if:

(i) for every $\boldsymbol{\theta} \in \Theta$, $\mathcal{O}(\boldsymbol{\theta}) = \mathcal{M}(\boldsymbol{\theta}; \boldsymbol{\theta})$ holds.

(ii) for every $\boldsymbol{\theta}, \boldsymbol{v} \in \Theta$ such that $\boldsymbol{\theta} \neq \boldsymbol{v}$, $\mathcal{O}(\boldsymbol{\theta}) \leq \mathcal{M}(\boldsymbol{\theta}; \boldsymbol{v})$ holds.

Let $\boldsymbol{\theta}^{(0)}$ be some initial value and $\boldsymbol{\theta}^{(r)}$ be a sequence of iterates (in $r$) for computing (10). Definition 1 suggests the following scheme that we will refer to as an MM algorithm.

**Definition 2.** Let $\boldsymbol{\theta}^{(0)}$ be some initial value and $\boldsymbol{\theta}^{(r)}$ be the $r$th iterate. We say that $\boldsymbol{\theta}^{(r+1)}$ is the $(r+1)$th iterate of an MM algorithm if it satisfies

$$\boldsymbol{\theta}^{(r+1)} = \arg\min_{\boldsymbol{\theta} \in \Theta} \mathcal{M}\left(\boldsymbol{\theta}; \boldsymbol{\theta}^{(r)}\right).$$

From Definitions 1 and 2, we can deduce the monotonicity property of all MM algorithms. That is, if $\boldsymbol{\theta}^{(r)}$ is an sequence of MM algorithm iterates, the objective sequence $\mathcal{O}\left(\boldsymbol{\theta}^{(r)}\right)$ is monotonically decreasing in $r$.

**Proposition 1.** *If $\mathcal{M}(\boldsymbol{\theta}; \boldsymbol{v})$ be a majorizer of the objective $\mathcal{O}(\boldsymbol{\theta})$ and $\boldsymbol{\theta}^{(r)}$ is a sequence of MM algorithm iterates, then*

$$\mathcal{O}\left(\boldsymbol{\theta}^{(r+1)}\right) \leq \mathcal{M}\left(\boldsymbol{\theta}^{(r+1)}; \boldsymbol{\theta}^{(r)}\right) \leq \mathcal{M}\left(\boldsymbol{\theta}^{(r)}; \boldsymbol{\theta}^{(r)}\right) = \mathcal{O}\left(\boldsymbol{\theta}^{(r)}\right). \quad (11)$$

*Remark* 1. It is notable that an algorithm need not be an MM algorithm in the strict sense of Definition 2 in order for (11) to hold. In fact, any algorithm where the $(r+1)$th iterate satisfies

$$\boldsymbol{\theta}^{(r+1)} \in \left\{\boldsymbol{\theta} \in \Theta : \mathcal{M}\left(\boldsymbol{\theta}; \boldsymbol{\theta}^{(r)}\right) \leq \mathcal{M}\left(\boldsymbol{\theta}^{(r)}; \boldsymbol{\theta}^{(r)}\right)\right\}$$

will generate a monotonically decreasing sequence of objective evaluates. Such an algorithm can be thought of as a generalized MM algorithm, analogous to the generalized EM algorithms of Dempster et al. (1977); see also McLachlan & Krishnan (2008, Sec. 3.3).

Starting from some initial value $\boldsymbol{\theta}^{(0)}$, let $\boldsymbol{\theta}^{(\infty)} = \lim_{r \to \infty} \boldsymbol{\theta}^{(r)}$ be the limit point of a sequence of algorithm iterates, if it exists. The following result from Razaviyayn et al. (2013) provides a strong statement regarding the global convergence of MM algorithm iterate sequences.

**Proposition 2.** *Starting from some initial value $\boldsymbol{\theta}^{(0)}$, if $\boldsymbol{\theta}^{(\infty)}$ is the limit point*

of an MM algorithm sequence of iterates $\boldsymbol{\theta}^{(r)}$ (i.e. satisfying Definition 2), then $\boldsymbol{\theta}^{(\infty)}$ is a stationary point of the problem (10).

*Remark* 2. Proposition 2 only guarantees the convergence of MM algorithm iterates to a stationary point and not a global, or even a local minimum. As such, for problems over difficult objective functions, multiple or good initial values are required in order to ensure that the obtained solution is of a high quality. Furthermore, Proposition 2 only guarantees convergence to a stationary point of (10) if a limit point exists for the chosen starting value. If a limit point does not exist then the MM algorithm objective sequence may diverge. This is a common problem in ML estimation of Gaussian mixture models [cf. McLachlan & Peel (2000, Sec. 3.9.1)].

## 3.1 Useful majorizers

There is an abundant literature on functions that can be used as majorizers and applications of such functions. Some early and fundamental works such as Becker et al. (1997), Bohning (1992), Bohning & Lindsay (1988), De Pierro (1993), and Heiser (1995) established many of the basic techniques for majorization. More modern majorizers for statistical and generic optimization problems can be found in Lange (2013), Lange (2016), and McLachlan & Krishnan (2008). We now present three majorizers that will be useful in constructing MM algorithms for the problems from Section 2.

**Fact 1.** *Let* $\mathcal{O}(\boldsymbol{\theta}) = \psi(\boldsymbol{a}^\top \boldsymbol{\theta})$ *where* $\psi : [0, \infty) \to \mathbb{R}$ *is a convex function. If* $\boldsymbol{a}, \boldsymbol{\theta}, \boldsymbol{v} \in [0, \infty)^d$ *for some* $d \in \mathbb{N}$, *then* $\mathcal{O}(\boldsymbol{\theta})$ *is majorized by*

$$\mathcal{M}(\boldsymbol{\theta}; \boldsymbol{v}) = \sum_{c=1}^{d} \frac{a_c v_c}{\boldsymbol{a}^\top \boldsymbol{v}} \psi\left(\frac{\boldsymbol{a}^\top \boldsymbol{v}}{v_c} \theta_c\right),$$

*where* $a_c, \theta_c, v_c$ *are the cth elements of* $\boldsymbol{a}, \boldsymbol{\theta}, \boldsymbol{v}$, *respectively, for* $c \in [d]$.

**Fact 2.** *Let $\mathcal{O}(\boldsymbol{\theta})$ be twice differentiable in $\boldsymbol{\theta}$. If $\boldsymbol{\theta}, \boldsymbol{v} \in \Theta \subset \mathbb{R}^d$ and $\mathbf{H} - \mathcal{H}_{\boldsymbol{\theta}}\mathcal{O}$ is positive semidefinite for all $\boldsymbol{\theta}$, then $\mathcal{O}(\boldsymbol{\theta})$ is majorized by*

$$\mathcal{M}(\boldsymbol{\theta}; \boldsymbol{v}) = \mathcal{O}(\boldsymbol{v}) + \nabla \mathcal{O}(\boldsymbol{v})(\boldsymbol{\theta} - \boldsymbol{v}) + \frac{1}{2}(\boldsymbol{\theta} - \boldsymbol{v})^\top \mathbf{H}(\boldsymbol{\theta} - \boldsymbol{v}).$$

**Fact 3.** *Let $d \in [1, 2]$ and let $\mathcal{O}(\theta) = |\theta|^d$. If $\theta, \upsilon \ne 0$, then $\mathcal{O}(\theta)$ is majorized by*

$$\mathcal{M}(\theta; \upsilon) = \frac{d}{2}|\upsilon|^{d-2}\theta^2 + \left(1 - \frac{d}{2}\right)|\upsilon|^d.$$

As an example, using Fact 3, the functions $\mathcal{O}_1(\theta) = |\theta|$ and $\mathcal{O}_{1.5}(\theta) = |\theta|^{1.5}$ can be majorized at $\upsilon = 1/2$ by $\mathcal{M}_1(\theta; 1/2) = \theta^2 + 1/4$ and $\mathcal{M}_{1.5}(\theta; 1/2) = (12\theta^2 + 1)/(8\sqrt{2})$, respectively. Plots of the example objectives and respective majorizers appear in Figure 1.

## 4 Examples of MM algorithms

### 4.1 Gaussian mixture of regressions

We use the notation from Section 2.1. Given the $r$th iterate of an MM algorithm, for each $i \in [n]$, set $\psi = -\log$, and let $\boldsymbol{a}^\top = (1, ..., 1)$,

$$\boldsymbol{\theta}^\top = (\pi_1 \phi(\boldsymbol{y}_i; \mathbf{B}_1 \boldsymbol{x}_i, \boldsymbol{\Sigma}_1), ..., \pi_g \phi(\boldsymbol{y}_i; \mathbf{B}_g \boldsymbol{x}_i, \boldsymbol{\Sigma}_g)),$$

and

$$\boldsymbol{v}^\top = \left(\pi_1 \phi\left(\boldsymbol{y}_i; \mathbf{B}_1^{(r)} \boldsymbol{x}_i, \boldsymbol{\Sigma}_1^{(r)}\right), ..., \pi_g \phi\left(\boldsymbol{y}_i; \mathbf{B}_g^{(r)} \boldsymbol{x}_i, \boldsymbol{\Sigma}_g^{(r)}\right)\right).$$

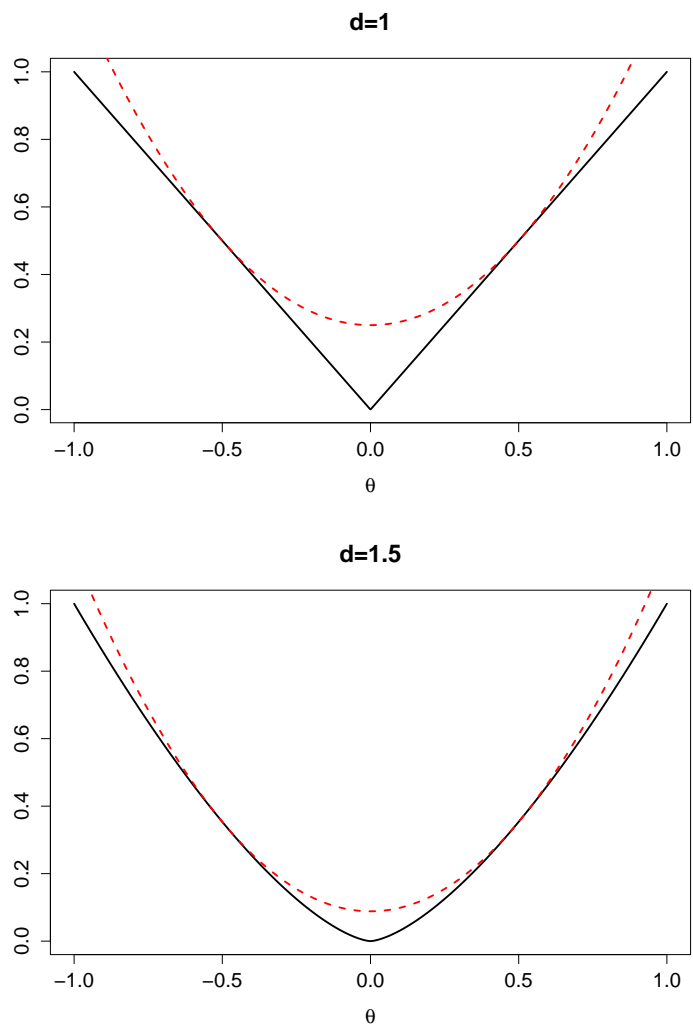

Figure 1: Examples of majorizers for objectives of the form $\mathcal{O}(\theta) = |\theta|^d$ at the point $v = 1/2$, for $d = 1, 1.5$. The solid lines indicate the objectives and the dashed lines indicate the majorizers.

Fact 1 suggests a majorizer for (5) of the form

$$\mathcal{M}\left(\boldsymbol{\theta};\boldsymbol{\theta}^{(r)}\right) = -\sum_{c=1}^{g}\sum_{i=1}^{n}\tau_{ci}^{(r)}\left[\log\pi_c + \log\phi\left(\boldsymbol{y}_i;\mathbf{B}_c\boldsymbol{x}_i,\boldsymbol{\Sigma}_c\right)\right]$$
$$+\sum_{c=1}^{g}\sum_{i=1}^{n}\tau_{ci}^{(r)}\log\left(\tau_{ci}^{(r)}\right), \quad (12)$$

where $\tau_{ci}^{(r)} = \pi_c\phi\left(\boldsymbol{y}_i;\mathbf{B}_c^{(r)}\boldsymbol{x}_i,\boldsymbol{\Sigma}_c^{(r)}\right)/\sum_{d=1}^{g}\pi_d\phi\left(\boldsymbol{y}_i;\mathbf{B}_d^{(r)}\boldsymbol{x}_i,\boldsymbol{\Sigma}_d^{(r)}\right)$, for each $c \in [g]$ and $i \in [n]$. Simplifying the first term of (12) via (3) yields

$$\mathcal{M}\left(\boldsymbol{\theta};\boldsymbol{\theta}^{(r)}\right) = -\sum_{c=1}^{g}\sum_{i=1}^{n}\tau_{ci}^{(r)}\log\pi_c + \frac{1}{2}\sum_{c=1}^{g}\sum_{i=1}^{n}\tau_i^{(r)}\log|\boldsymbol{\Sigma}_c|$$
$$+\frac{1}{2}\sum_{c=1}^{g}\sum_{i=1}^{n}\tau_{ci}^{(r)}\left(\boldsymbol{y}_i - \mathbf{B}_c\boldsymbol{x}_i\right)^{\top}\boldsymbol{\Sigma}^{-1}\left(\boldsymbol{y}_i - \mathbf{B}_c\boldsymbol{x}_i\right) \quad (13)$$
$$+\mathcal{C}^{(r)},$$

where $\mathcal{C}^{(r)}$ is a constant that does not depend on the parameter $\boldsymbol{\theta}$.

Under the restrictions on $\pi_c$, we must minimize (13) over the constrained parameter space

$$\Theta = \left\{\boldsymbol{\theta} : \pi_c > 0, \sum_{c=1}^{g}\pi_c = 1, \mathbf{B}_c \in \mathbb{R}^{q\times p}, \boldsymbol{\Sigma}_c \text{ is positive definite}, c \in [g]\right\}.$$

This can be achieved by computing the roots of $\nabla_{\boldsymbol{\theta}}\Lambda = \mathbf{0}$, where

$$\Lambda\left(\boldsymbol{\theta};\boldsymbol{\theta}^{(r)}\right) = \mathcal{M}\left(\boldsymbol{\theta};\boldsymbol{\theta}^{(r)}\right) + \lambda\left(\sum_{c=1}^{g}\pi_c - 1\right)$$

is the Lagrangian with multiplier $\lambda \in \mathbb{R}$. The resulting $(r+1)$th iterate of the MM algorithm for the ML estimation of the GMR model can be defined as $\boldsymbol{\theta}^{(r)}$, which contains the elements

$$\pi_c^{(r+1)} = n^{-1} \sum_{i=1}^{n} \tau_{ci}^{(r)}, \tag{14}$$

$$\mathbf{B}_c^{(r+1)} = \left(\sum_{i=1}^{n} \tau_{ci}^{(r)} \boldsymbol{y}_i \boldsymbol{x}_i^\top\right) \left(\sum_{i=1}^{n} \tau_{ci}^{(r)} \boldsymbol{x}_i \boldsymbol{x}_i^\top\right)^{-1}, \tag{15}$$

and

$$\boldsymbol{\Sigma}_c = \frac{\sum_{i=1}^{n} \tau_{ci}^{(r)} \left(\boldsymbol{y}_i - \mathbf{B}_c^{(r+1)} \boldsymbol{x}_i\right) \left(\boldsymbol{y}_i - \mathbf{B}_c^{(r+1)} \boldsymbol{x}_i\right)^\top}{\sum_{i=1}^{n} \tau_{ci}^{(r)}}. \tag{16}$$

*Remark* 3. The MM algorithm defined by updates (14)–(16) is exactly the same as the EM algorithm for ML estimation that is derived in Jones & McLachlan (1992). There are numerous cases where MM and EM algorithms coincide and some conditions under which such coincidences occur are explored in Meng (2000).

*Remark* 4. Note that updates (15) and (16) require matrix additions, multiplications, and inversions that may be computationally prohibitive if $n$, $p$, and $q$ are large. It is possible to modify the MM algorithm via the techniques from Nguyen & McLachlan (2015) to avoid such matrix computations. Such modifications come at a cost of slower convergence of the algorithm, but can make GMR feasible for data sets that were prohibitively large without such changes.

### 4.2 Multinomial-logistic regressions

We use the notation from Section 2.2. Consider only the $c$th set of parameter elements $\boldsymbol{\beta}_c$. The gradient and second-order derivatives of $\mathcal{R}$ with respect to $\boldsymbol{\beta}_c$ can be written as

$$\nabla_{\boldsymbol{\beta}_c} \mathcal{R} = -\frac{1}{n} \sum_{i=1}^{n} \left[ \mathbb{I}\{y_i = c\} - \frac{\exp\left(\boldsymbol{\beta}_c^\top \boldsymbol{x}_i\right)}{\sum_{d=1}^{g} \exp\left(\boldsymbol{\beta}_d^\top \boldsymbol{x}_i\right)} \right] \boldsymbol{x}_i$$

and

$$\nabla_{\boldsymbol{\beta}_c}\nabla_{\boldsymbol{\beta}_c^\top}\mathcal{R} = \frac{1}{n}\sum_{i=1}^{n}\frac{\exp\left(\boldsymbol{\beta}_c^\top\boldsymbol{x}_i\right)}{\sum_{d=1}^{g}\exp\left(\boldsymbol{\beta}_d^\top\boldsymbol{x}_i\right)}\left[1-\frac{\exp\left(\boldsymbol{\beta}_c^\top\boldsymbol{x}_i\right)}{\sum_{d=1}^{g}\exp\left(\boldsymbol{\beta}_d^\top\boldsymbol{x}_i\right)}\right]\boldsymbol{x}_i\boldsymbol{x}_i^\top.$$

Define $\Pi = \exp\left(\boldsymbol{\beta}_c^\top\boldsymbol{x}\right)/\sum_{d=1}^{g}\exp\left(\boldsymbol{\beta}_d^\top\boldsymbol{x}\right)$; it can be shown that $\Pi\left(1-\Pi\right) \leq 1/4$ via calculus. Thus, we have the fact that $\Delta/4 - \nabla_{\boldsymbol{\beta}_c}\nabla_{\boldsymbol{\beta}_c^\top}\mathcal{R}$ is positive definite, since $\Delta = n^{-1}\sum_{i=1}^{n}\boldsymbol{x}_i\boldsymbol{x}_i^\top$ is positive definite except for in pathological cases. Let $\boldsymbol{\theta}^{(r)}$ be the $r$th iterate of the MM algorithm, Fact 1 yields the majorizers

$$\begin{aligned}\mathcal{M}_c\left(\boldsymbol{\beta}_c;\boldsymbol{\theta}^{(r)}\right) &= \mathcal{R}\left(\boldsymbol{\theta}^{(r)};\bar{\mathbb{D}}\right) + \nabla_{\boldsymbol{\beta}_c}\mathcal{R}_{\boldsymbol{\theta}^{(r)}}\left(\boldsymbol{\beta}_c - \boldsymbol{\beta}_c^{(r)}\right)\\ &\quad + \frac{1}{8}\left(\boldsymbol{\beta}_c - \boldsymbol{\beta}_c^{(r)}\right)^\top\Delta\left(\boldsymbol{\beta}_c - \boldsymbol{\beta}_c^{(r)}\right),\end{aligned} \quad (17)$$

for each $c \in [g-1]$, by setting $\mathbf{H} = \Delta/4$. Here, $\nabla_{\boldsymbol{\beta}_c}\mathcal{R}_{\boldsymbol{\theta}^{(r)}}$ is the gradient with respect to $\boldsymbol{\beta}_c$, with $\boldsymbol{\theta}$ evaluated at $\boldsymbol{\theta}^{(r)}$.

Given $\boldsymbol{\theta}^{(r)}$, $\mathcal{M}_c\left(\boldsymbol{\beta}_c;\boldsymbol{\theta}^{(r)}\right)$ can be globally minimized by solving the FOC equation $\nabla_{\boldsymbol{\beta}_c}\mathcal{M}_c = \mathbf{0}$, which yields the solution

$$\boldsymbol{\beta}_c^* = \boldsymbol{\beta}_c^{(r)} + 4\Delta^{-1}\nabla_{\boldsymbol{\beta}_c}\mathcal{R}_{\boldsymbol{\theta}^{(r)}}. \quad (18)$$

Since only the $c$th parameter element is majorized by (17), the solution (18) suggests the following algorithm: let $\boldsymbol{\theta}^{(r)}$ be the $r$th iterate of the algorithm, on the $(r+1)$th iteration, set $\boldsymbol{\theta}^{(r+1)\top} = \left(\boldsymbol{\beta}_1^{(r+1)\top},...,\boldsymbol{\beta}_{g-1}^{(r+1)\top}\right)$ according to the update rule

$$\boldsymbol{\beta}_c^{(r+1)} = \begin{cases}\boldsymbol{\beta}_c^* & \text{if } c = 1 + (r \bmod g - 1),\\ \boldsymbol{\beta}_c^{(r)} & \text{otherwise.}\end{cases} \quad (19)$$

*Remark* 5. The algorithm as defined by update rule (18) is an example of a generalized MM algorithm as discussed in Remark 1 and thus satisfies inequality

(11) but does not satisfy the conditions for Proposition 2. To show the global convergence of rule (18), we can note that it is an example of a block SUM algorithm and demonstrate the satisfaction of the assumptions for Razaviyayn et al. (2013, Thm. 2).

*Remark* 6. The update rule (18) allows for blockwise update of the parameter elements $\boldsymbol{\beta}_c$ rather than all at once, as is required via a Newton-Raphson approach. Furthermore, the only large matrix computation that is required is the matrix inversion of $\Delta$, which is only required to be conducted once as it does not depend on the iterates $\boldsymbol{\theta}^{(r)}$.

### 4.3 Support vector machines

We use the notation from Section 2.3. Consider the identity

$$\max\{a, b\} = \frac{1}{2}|a - b| + \frac{a+b}{2} \tag{20}$$

for $a, b \in \mathbb{R}$. Using (20) and Fact 3, we can derive the majorizer

$$\mathcal{M}(\theta; \upsilon) = \frac{1}{4|\upsilon|}(\theta + |\upsilon|)^2,$$

for the objective $\mathcal{O}(\theta) = \max\{0, \theta\}$, at $\upsilon \neq 0$. To avoid the singularity at $\upsilon = 0$, de Leeuw & Lange (2009) suggests the approximate majorizer

$$\mathcal{M}_\epsilon(\theta; \upsilon) = \frac{1}{4|\upsilon| + \epsilon}(\theta + |\upsilon|)^2, \tag{21}$$

where $\epsilon = 10^{-5}$ is sufficiently small for double-precision computing. Using (21), we can majorize (9) at the $r$th algorithm iterate by making the substitutions $\theta = 1 - y_i(\alpha + \boldsymbol{\beta}^\top \boldsymbol{x}_i)$ and $\upsilon = 1 - y_i(\alpha^{(r)} + \boldsymbol{\beta}^{(r)\top}\boldsymbol{x}_i)$, for each $i \in [n]$. The resulting majorizer of $n$ times the risk (9) at $\boldsymbol{\theta}^{(r)}$ has the form

$$\mathcal{M}_\epsilon\left(\boldsymbol{\theta};\boldsymbol{\theta}^{(r)}\right) = \sum_{i=1}^{n} \frac{1}{4w_i^{(r)} + \epsilon} \left[w_i^{(r)} + 1 - y_i\left(\alpha + \boldsymbol{\beta}^\top \boldsymbol{x}_i\right)\right]^2 + n\lambda\boldsymbol{\beta}^\top\boldsymbol{\beta}, \quad (22)$$

where $w_i^{(r)} = \left|1 - y_i\left(\alpha^{(r)} + \boldsymbol{\beta}^{(r)\top}\boldsymbol{x}_i\right)\right|$, for each $i$. Write $\tilde{\boldsymbol{x}}_i^\top = \left(y_i, y_i\boldsymbol{x}_i^\top\right)$ and $\tilde{w}_i^{(r)} = w_i^{(r)} + 1$ for each $i$, and let $\tilde{\mathbf{I}} = \mathrm{diag}\left(0, 1, ..., 1\right)$ and

$$\boldsymbol{\Omega}^{(r)} = \mathrm{diag}\left(\frac{1}{4w_1^{(r)} + \epsilon}, ..., \frac{1}{4w_n^{(r)} + \epsilon}\right).$$

Put $\tilde{\boldsymbol{x}}_i$ in the $i$th row of the matrix $\mathbf{X} \in \mathbb{R}^{n \times (p+1)}$, and set $\boldsymbol{w}^{(r)\top} = \left(\tilde{w}_1^{(r)}, ..., \tilde{w}_n^{(r)}\right)$. We can write (22) in matrix form as

$$\mathcal{M}\left(\boldsymbol{\theta};\boldsymbol{\theta}^{(r)}\right) = \left(\boldsymbol{w}^{(r)} - \mathbf{X}\boldsymbol{\theta}\right)^\top \boldsymbol{\Omega}^{(r)} \left(\boldsymbol{w}^{(r)} - \mathbf{X}\boldsymbol{\theta}\right) + n\lambda\boldsymbol{\theta}^\top\tilde{\mathbf{I}}\boldsymbol{\theta}. \quad (23)$$

Majorizer (23) is in quadratic form and thus has the global minimum solution

$$\boldsymbol{\theta}^{(r+1)} = \left(\mathbf{X}^\top\boldsymbol{\Omega}^{(r)}\mathbf{X} + n\lambda\tilde{\mathbf{I}}\right)^{-1}\mathbf{X}^\top\boldsymbol{\Omega}^{(r)}\boldsymbol{w}^{(r)}. \quad (24)$$

The MM algorithm for the linear-basis SVM can thus be defined via the IRLS rule (24).

*Remark* 7. The MM algorithm defined via (24) is similar to the IRLS algorithm of Navia-Vasquez et al. (2001). However, whereas our weightings are obtained via simple majorization, the weightings used by Navia-Vasquez et al. (2001) are obtained via careful manipulation of the KKT conditions.

# 5 Numerical demonstrations

## 5.1 Gaussian mixture of regressions

There are numerous packages in the **R** (R Core Team, 2016) programming language that implement the EM/MM algorithm for estimating GMR models; see Remark 3. These packages include **EMMIXcontrasts** (Ng et al., 2014), **flexmix** (Grun & Leisch, 2008), and **mixtools** (Benaglia et al., 2009).

Using **R**, we simulate data according to Case 2 of the sampling experiments of Quandt (1972). That is, we simulate $n = 120$ observations, where $\boldsymbol{X}_i^\top = (1, U_i)$ with $U_i$ uniformly distributed between 10 and 20, for $i \in [n]$. Conditioned on $\boldsymbol{X}_i = \boldsymbol{x}_i$, we simulate $Y_i$ according to the two-component GMR model

$$f_{Y|X}(y_i|\boldsymbol{x}_i; \boldsymbol{\theta}) = \frac{1}{2}\phi(y_i; (1,1)\,\boldsymbol{x}_i, 2) + \frac{1}{2}\phi(y_i; (0.5, 1.5)\,\boldsymbol{x}_i, 2.5).$$

In Table 2, we present the negative log-likelihood risk (5) values for 20 iterations of the EM/MM algorithm to estimate a $g = 2$ component GMR using the **regmixEM** function from **mixtools**, for a single instance of the experimental setup. We see that the risk values are monotonically decreasing in accordance with Proposition 1. Figure 2 display the simulated data, the generative mean model, and the fitted conditional mean functions $\hat{y}_c(\boldsymbol{x}) = \left(\hat{b}_{c1}, \hat{b}_{c2}\right)\boldsymbol{x}$, for each of the fitted GMR components $c = 1, 2$. Here the estimate of the parameter elements $\hat{b}_{c1}, \hat{b}_{c2}$ are contained within $\hat{\boldsymbol{\theta}}$.

## 5.2 Multinomial-logistic regressions

With **R,** we utilize algorithm (19) to compute (1) for the problem of estimating an MLR for the Fisher's Iris data set (Fisher, 1936). The data is a part of the core **datasets** package of **R** and contains $n = 150$ observations, 50 each, from

Table 2: Negative log-likelihood risk for the estimation of a $g = 2$ component GMR model via an EM/MM algorithm.

| Iteration | Risk     | Iteration | Risk     |
|-----------|----------|-----------|----------|
| 1         | 325.8761 | 11        | 294.5583 |
| 2         | 317.1869 | 12        | 294.5501 |
| 3         | 314.0193 | 13        | 294.5477 |
| 4         | 311.5091 | 14        | 294.5470 |
| 5         | 308.0247 | 15        | 294.5468 |
| 6         | 302.8068 | 16        | 294.5467 |
| 7         | 297.5655 | 17        | 294.5467 |
| 8         | 295.1926 | 18        | 294.5467 |
| 9         | 294.6909 | 19        | 294.5467 |
| 10        | 294.5865 | 20        | 294.5467 |

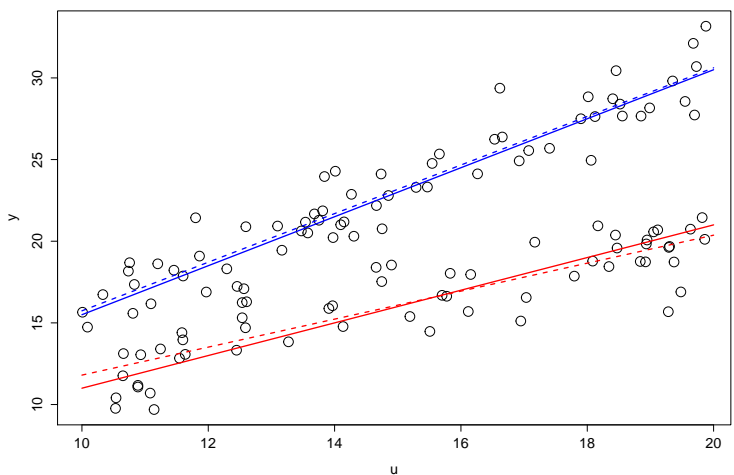

Figure 2: Example of an instance of the Case 2 of the sampling experiments of Quandt (1972). Solid lines indicate generative mean functions according to the $g = 2$ different components of the mixture, and dashed lines indicate fitted mean functions $\hat{y}_c(\boldsymbol{x})$ for $c = 1, 2$.

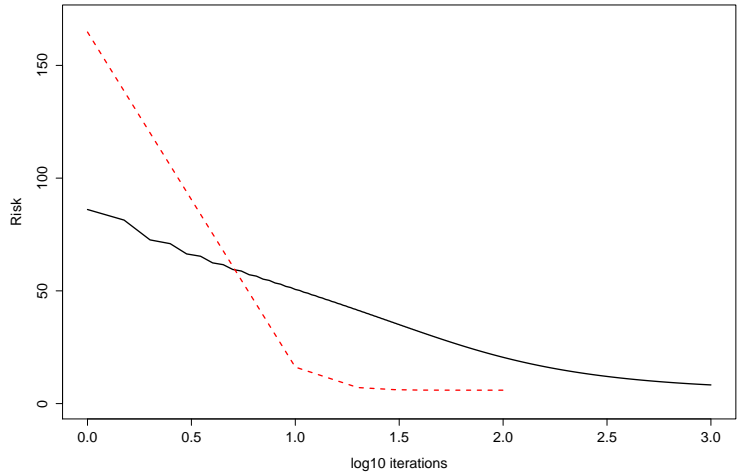

Figure 3: Negative log-likelihood risk versus $\log_{10}$ iterations for the MLR model fitted to the **iris** data set. The solid line indicates the MM algorithm-obtained sequence, and the dashed line indicates the sequence obtained from the **multinom** function.

$g = 3$ species of irises. Each observation consists of a feature vector

$$\boldsymbol{u}_i^\top = (\text{Sepal Length}_i,\ \text{Sepal Width}_i,\ \text{Petal Length}_i,\ \text{Petal Width}_i),$$

along with a species name $y_i \in \{\text{Setosa, Versicolor, Virginica}\}$, for $i \in [n]$. We map the species names to the set $[g]$ for convenience and set $\boldsymbol{x}_i^\top = \left(1, \boldsymbol{u}_i^\top\right)$.

A plot of the negative log-likelihood risk versus the logarithm of the number of iterations is presented in Figure 3, along with the risk sequence obtained from the **multinom** function of the **nnet** package (Venables & Ripley, 2002), which solves the same problem. The difference in convergence speed between the two algorithms is not surprising as **multinom** utilizes a Newton-Raphson algorithm, which exhibits quadratic-rate convergence to stationary points within a close enough radius to the limit point, whereas MM algorithms only exhibit linear-rate convergence [cf. Lange (2013, Ch. 12)].

*Remark* 8. Some practical suggestions for acceleration of convergence speed for MM algorithms are provided in Lange (2016, Ch. 7). The simplest of such suggestions is to simply double each MM iterate. That is, if at the $(r+1)$ th step, we make the update $\boldsymbol{\theta}^{(r+1)} = \mathcal{U}\left(\boldsymbol{\theta}^{(r)}\right)$, then we instead make the update

$$\boldsymbol{\theta}^{(r+1)} = \boldsymbol{\theta}^{(r)} + 2\left[\mathcal{U}\left(\boldsymbol{\theta}^{(r)}\right) - \boldsymbol{\theta}^{(r)}\right].$$

At most, the new updates can halve the number of iterations required. However, fulfillment of inequality (11) is no longer guaranteed.

### 5.3 Support vector machines

For ease of visualization, we now concentrate on the first $n = 100$ observations from the **iris** data, at the two input variables

$$\boldsymbol{x}_i^\top = (\text{Sepal Length}_i, \ \text{Sepal Width}_i)$$

for $i \in [n]$. The species names, Setosa and Versicolor, are mapped to $-1$ and $1$, respectively, and thus $y_i \in \{-1, 1\}$ depending on the species of observation $i$. A linear-basis SVM is fitted using algorithm (24) with $\lambda = 0.1$ for 30 iterations. Table 3 presents the risk (9) at each IRLS/MM iteration, and Figure 4 displays the resulting seperating hyperplane.

From Table 3, we note that the IRLS/MM algorithm convergences quickly in the SVM problem and that the risk is monotonically decreasing as expected. Further, we note that for this subset of the **iris** data, the separating hyperplane can perfectly separate the two classes; this is not always possible in general.

Table 3: SVM risk for the **iris** data with $\lambda = 0.1$, obtained via the IRLS/MM algorithm.

| Iteration | Risk | Iteration | Risk |
|---|---|---|---|
| 1 | 47.36808 | 16 | 47.20901 |
| 2 | 47.29061 | 17 | 47.20895 |
| 3 | 47.26262 | 18 | 47.20891 |
| 4 | 47.25164 | 19 | 47.20888 |
| 5 | 47.24096 | 20 | 47.20886 |
| 6 | 47.23068 | 21 | 47.20885 |
| 7 | 47.22281 | 22 | 47.20884 |
| 8 | 47.21746 | 23 | 47.20884 |
| 9 | 47.21405 | 24 | 47.20883 |
| 10 | 47.21193 | 25 | 47.20883 |
| 11 | 47.21064 | 26 | 47.20883 |
| 12 | 47.20988 | 27 | 47.20883 |
| 13 | 47.20946 | 28 | 47.20882 |
| 14 | 47.20923 | 29 | 47.20882 |
| 15 | 47.20910 | 30 | 47.20882 |

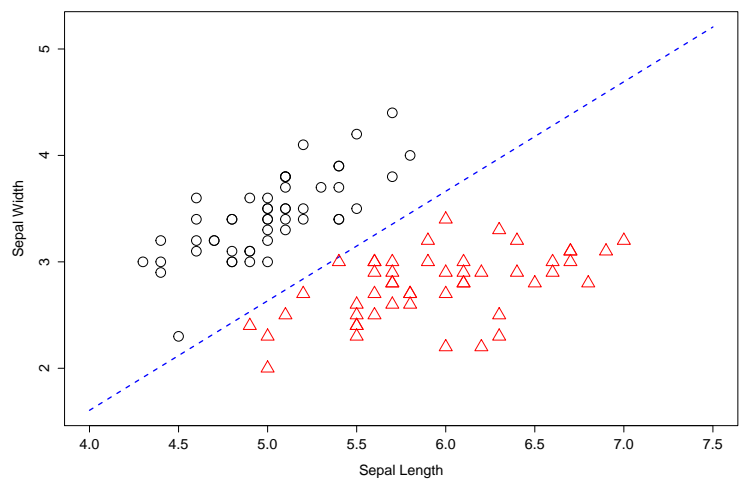

Figure 4: Separating hyperplane between for the classification problem of separating the Setosa and Versicolor irises by their sepal length and width. The dashed line indicates the SVM-obtained separating hyperplane. The Circles and Triangles indicate Setosa and Versicolor irises, respectively.

# 6 Conclusions

The MM algorithm framework is a popular tool for deriving useful algorithms for problems in machine learning and statistical estimation. We have introduced and demonstrated its usage in three example applications: Gaussian mixture regressions, multinomial logistic regressions, and support vector machines; which are commonly applied by practitioners for solving real-world problems. Although example-based, the techniques that are introduced in this article are by no means restricted to the specific examples, nor even to machine learning and statistical estimation; see Chi et al. (2014) and Lange & Zhou (2014) for example.

We note that there are aspects of the MM algorithm framework that we have omitted for brevity. For example, we have not discussed the manner in which to choose to terminate an MM algorithm, as this is often a contentious point. A good discussion on the relative merits of different methods can be found in Lange (2013, Sec. 11.5). Additionally, we have not discussed the many computational benefits of MM algorithms, such as parallelizability and distributability. A case study of parallelizability for heteroscedastic regression is provided in Nguyen et al. (2016a). General discussions regarding the implementation of MM algorithms on graphical processing units, in parallel, are provided in Zhou et al. (2010).

It is hoped that this article demonstrates the usefulness and ubiquity of the MM algorithm framework to the reader. For enthusiastic and interested readers, we highly recommend the outstanding treatment of the topic in Lange (2016).

## Acknowledgment

The author is grateful to Prof. Geoffrey McLachlan for his constant encouragement and support. The author thanks the ARC for their financial support.

Meng, X.-L. (2000). Optimization transfer using surrogate objective functions: Discussion. *Journal of Computational and Graphical Statistics*, 9, 35–43.

Navia-Vasquez, A., Perez-Cruz, F., Artes-Rodriguez, A., & Figueiras-Vidal, A. R. (2001). Weighted least squares training of support vector classifiers leading to compact and adaptive schemes. *IEEE Transactions on Neural Networks*, 12, 1047–1059.

Ng, S.-K., McLachlan, G., & Wang, K. (2014). *EMMIXcontrasts: Contrasts in mixed effects for EMMIX model with random effects*.

Nguyen, H. D., Lloyd-Jones, L. R., & McLachlan, G. J. (2016a). A block minorization-maximization algorithm for heteroscedastic regression. *IEEE Signal Processing Letters*, 23, 1031–1135.

Nguyen, H. D. & McLachlan, G. J. (2014). Asymptotic inference for hidden process regression models. In *Proceedings of the 2014 IEEE Statistical Signal Processing Workshop*.

Nguyen, H. D. & McLachlan, G. J. (2015). Maximum likelihood estimation of Gaussian mixture models without matrix operations. *Advances in Data Analysis and Classification*, 9, 371–394.

Nguyen, H. D. & McLachlan, G. J. (2016a). Laplace mixture of linear experts. *Computational Statistics and Data Analysis*, 93, 177–191.

Nguyen, H. D. & McLachlan, G. J. (2016b). Maximum likelihood estimation of triangular and polygonal distributions. *Computational Statistics and Data Analysis*, 106, 23–36.

Nguyen, H. D., McLachlan, G. J., Ullmann, J. F. P., & Janke, A. L. (2016b). Laplace mixture autoregressive models. *Statistics and Probability Letters*, 110, 18–24.30